\documentclass[12pt]{iopart}
\usepackage{graphicx}

\def\gtwid{\mathrel{\raise.3ex\hbox{$>$\kern-.75em\lower1ex\hbox{$\sim$}}}}
\def\alt{\mathrel{\raise.3ex\hbox{$<$\kern-.75em\lower1ex\hbox{$\sim$}}}}
\def\agt{\mathrel{\raise.3ex\hbox{$>$\kern-.75em\lower1ex\hbox{$\sim$}}}}

\newcommand{\be}{\begin{equation}}
\newcommand{\ee}{\end{equation}}

\newcommand\Rey{\mbox{\textrm{Re}}}  

\newcommand\Pra{\mbox{\textrm{Pr}}}  
\newcommand\Ra{\mbox{\textrm{Ra}}}  
\newcommand\Nu{\mbox{\textrm{Nu}}}  

\begin{document}

\title[Heat transport by turbulent convection for $\Gamma = 1.00$ and $\Ra\ \alt 2\times 10^{14}$]{Heat transport by turbulent Rayleigh-B\'enard convection for $\Pra\ \simeq 0.8$ and $4\times 10^{11} \alt \Ra\ \alt 2\times10^{14}$: Ultimate-state transition for aspect ratio $\Gamma = 1.00$}

\author{Xiaozhou He$^{1,6}$, Denis Funfschilling$^{2,6}$, Eberhard Bodenschatz$^{1,3,4,6}, and $ Guenter Ahlers$^{5,6}$}

\address{$^{1}$Max Planck Institute for Dynamics and Self-Organization (MPIDS), 37077 G\"ottingen, Germany}
\address{$^2$LSGC CNRS - GROUPE ENSIC, BP 451, 54001 Nancy Cedex, France}
\address{$^{3}$Institute for Nonlinear Dynamics, University of G\"ottingen, 37077 G\"ottingen, Germany}
\address{$^{4}$Laboratory of Atomic and Solid-State Physics and Sibley School of Mechanical and Aerospace Engineering, Cornell University, Ithaca, NY 14853, USA}
\address{$^5$Department of Physics, University of California, Santa Barbara, CA 93106, USA}
\address{$^6$International Collaboration for Turbulence Research}

\begin{abstract}
We report experimental results for heat-transport measurements, in the form of the Nusselt number \Nu, by turbulent Rayleigh-B\'enard convection in a cylindrical sample of aspect ratio $\Gamma \equiv D/L = 1.00$ ($D = 1.12$ m is the diameter and $L = 1.12$ m the height) and compare them with previously reported results for $\Gamma = 0.50$. The measurements were made using sulfur hexafluoride at pressures up to 19 bars as the fluid. They are for the Rayleigh-number range $4\times10^{11} \alt \Ra \alt 2\times10^{14}$ and for Prandtl numbers \Pra\ between 0.79 and 0.86. 

For $\Ra < \Ra^*_1 \simeq 2\times 10^{13}$ we find $\Nu = N_0 \Ra^{\gamma_{eff}}$ with $\gamma_{eff} = 0.321 \pm 0.002$ and $N_0 = 0.0776$, consistent with classical turbulent Rayleigh-B\'enard convection in a system with laminar boundary layers below the top and above the bottom plate and with the prediction of Grossmann and Lohse. 

For $\Ra > \Ra_1^*$ the data rise above the classical-state power-law and show greater scatter. In analogy to similar behavior observed for  $\Gamma = 0.50$, we interpret this observation as the onset of the transition to the ultimate state. Within our resolution this onset occurs at  nearly the same value  of $\Ra_1^*$ as it does for $\Gamma = 0.50$. This differs from an earlier estimate by Roche {\it et al.} which yielded a transition at $\Ra_U \simeq 1.3\times 10^{11} \Gamma^{-2.5\pm 0.5}$.   A $\Gamma$-independent $\Ra^*_1$ would suggest that the boundary-layer shear transition is induced by fluctuations on a scale less than the sample dimensions rather than by a global $\Gamma$-dependent flow mode.  Within the  resolution of the measurements the heat transport above $\Ra_1^*$ is equal for the two $\Gamma$ values, suggesting a universal aspect of the ultimate-state transition and properties. The enhanced scatter of \Nu\ in the transition region, which exceeds the experimental resolution, indicates an intrinsic irreproducibility of the state of the system.

Several previous measurements for $\Gamma = 1.00$ are re-examined and compared with the present results. None of them identified the ultimate-state transition.

\end{abstract}

\maketitle

\section{Introduction}

In this paper we consider turbulent convection in a fluid contained between horizontal parallel plates and heated from below (Rayleigh-B\'enard convection or RBC; for reviews written for broad audiences see Refs.~\cite{Ka01,Ah09}; for more specialized reviews see Refs.~\cite{AGL09,LX10}). It is now well established experimentally that RBC for Rayleigh numbers \Ra\ (a dimensionless measure of the applied temperature difference $\Delta T$) below a typical value $\Ra^*_1$ is a system with laminar (albeit fluctuating) boundary layers \cite{ZX10,SZGVXL12} (BLs), one below the top and another above the bottom plate. Approximately half of $\Delta T \equiv T_b - T_t$ ($T_b$ and $T_t$ are the temperatures of the bottom and top confining plate respectively) is sustained by each of these BLs \cite{TBL93,BTL93,BTL94,XX97,LX98,ZX01,WX04,PRTBT07}). The sample interior, known as the ``bulk", is nearly isothermal in the time average  (see, however, Ref.~\cite{TBL93,BA07_EPL,WA11a,ABFGHLSV12}), but its temperature and velocity fields are also fluctuating vigorously. This state is known as the ``classical" state as it has been studied at great length for nearly a century.  

At very large \Ra\ a transition was predicted to take place \cite{Kr62,Sp71,GL11} from the classical state to the ``ultimate" state \cite{CCCHCC97} where the BLs have become turbulent as well because of the shear applied to them by the vigorous fluctuations in the sample interior. Experimentally it was found recently for a cylindrical sample with aspect ratio $\Gamma \equiv D/L = 0.50$ ($D$ is the diameter and $L$ the height of the cylindrical sample) and $\Pra \simeq 0.8$ that this transition takes place over a wide range $\Ra_1^* \alt \Ra \alt Ra_2^*$, with $\Ra_1^* \simeq 1.5\times 10^{13}$ and $\Ra_2^* \simeq 5\times 10^{14}$. For a more detailed description of the classical and ultimate state and the transition between them, see for instance Ref.~\cite{AHFB12} and the review articles \cite{Ka01,Ah09,AGL09}.

The purpose of the present work was two-fold. First we hoped to determine with high accuracy the dependence of \Nu\ on \Ra\ in the classical state at the largest-possible Rayleigh numbers for a sample of aspect ration $\Gamma = 1.00$ and for a Prandtl number $\Pra \simeq 0.8$. Such data make it possible to test  in detail the predictions for the classical state by Grossmann and Lohse \cite{GL00,GL01} of the relationship between \Nu\ and \Ra\  in a parameter range not explored heretofore. Although in principle these predictions should be applicable to the classical state regardless of $\Gamma$, they depend on a number of parameters that had been determined by fitting to experimental data for $\Gamma = 1.00$ \cite{AX01}. This fit was done over the range $4 \alt \Pra \alt 34$ and $3\times 10^7 \alt \Ra \alt 3\times 10^9$. Thus, a comparison with new data over the very different \Ra\ and \Pra\ ranges of the present work constitutes a significant test of the prediction. We found that $\Nu = N_0 \Ra^{\gamma_{eff}}$ with $N_0 = 0.0776$ and $\gamma_{eff} = 0.321 \pm 0.002$. This result differs slightly from the case $\Gamma = 0.50$ \cite{AHFB12} which yielded $\gamma_{eff} = 0.312 \pm 0.002$. It is in excellent agreement with the Grossmann-Lohse prediction for the classical state and $\Gamma = 1.00$ in our \Ra\ and \Pra\ range. 

Second, we hoped to search for the transition to the ``ultimate" state of turbulent convection. Experiments searching for this state using $\Gamma = 0.50$ had been carried out before \cite{CGHKLTWZZ89,CCCCCH96,CCCHCC97,NSSD00,NSSD00e,CCCCH01,RCCH01,RGCH05,GR08,GSBGPTR09,RGKS10}; results from these searches were reported and/or reviewed in another publication \cite{AHFB12}.  
The transition was found very recently \cite{HFNBA12,AHFB12} to occur over a wide \Ra-range, extending from $\Ra_1^* \simeq 2\times 10^{13}$ to $\Ra_2^* \simeq 5\times 10^{14}$. In the present project we focus on the particular case of a cylindrical sample with $\Gamma = 1.00$ ($D = 1.12$ m and $L=1.12$ m). This geometry was used in some previous searches for this state \cite{FG02,NS03,NS06a,RGKS10,UMS11} 
and thus enables a direct comparison with earlier measurements;  but more importantly we chose $\Gamma = 1.00$ in order to search for any $\Gamma$-dependence of the transition.  Earlier a transition in $\Nu(\Ra)$ had been reported at several $\Gamma$ values by Roche {\it et al.} \cite{RGKS10} at  Rayleigh numbers $\Ra_U \simeq 1.3\times 10^{11} \Gamma^{-2.5\pm 0.5}$ which those authors attributed to the ultimate-state transition. In contradistinction to this result, we find that the transition occurs at values of \Ra\ that are two orders of magnitude larger than $\Ra_U$, and that (for $\Gamma = 0.50$ and 1.00) $\Ra_1^*$ is independent of $\Gamma$ within the resolution of the data. 
A $\Gamma$-independent $\Ra^*_1$ would suggest that the boundary-layer shear-transition is induced by fluctuations on a scale less than the sample dimensions rather than by a global $\Gamma$-dependent flow mode.  Within the resolution of the results the heat transport above $\Ra_1^*$ is equal for the two $\Gamma$ values, suggesting a universal aspect of the ultimate-state transition and properties. Unfortunately the necessarily smaller height of the $\Gamma = 1.00$ sample (compared to $\Gamma = 0.50$) limited our measurement range to $\Ra \alt 2\times 10^{14}$ and prevented us from obtaining data all the way beyond $\Ra_2^*$.

Our results were obtained using the High-Pressure Convection Facility (the HPCF, a cylindrical sample of 1.12 m diameter) at the Max Planck Institute for Dynamics and Self-organization in G\"ottingen, Germany with sulfur hexafluoride (SF$_6$) at pressures up to 19 bars as the fluid. Results for $\Gamma = 0.50$ from this work were presented in Refs.~\cite{AFB09, AFB11a,AFB11b,HFNBA12,AHFB12}. A description of the apparatus was given in Ref.~\cite{AFB09}. The present paper presents new results obtained for a sample chamber known as HPCF-IV which had a height equal to its diameter.

In Sec.~\ref{sec:sys_paras} we define the parameters that describe this system. Then, in Sec.~\ref{sec:apparatus}, we give a brief discussion of  the apparatus used in this work. A detailed description of the main features was presented before \cite{AFB09}. Section~\ref{sec:Results} presents a comprehensive discussion of our results and of the results of others at large \Ra\ for cylindrical samples with $\Gamma = 1.00$. We conclude with a Summary in Sec.~\ref{sec:summary}.

\section{The system parameters and data analysis.}
\label{sec:sys_paras}

For turbulent RBC in cylindrical containers there are two parameters which, in addition to $\Gamma$, are expected to determine its state. They are the dimensionless temperature difference as expressed by the Rayleigh number
\be
\Ra \equiv \frac{\alpha g \Delta T L^3}{\kappa \nu}
\label{eq:Ra}
\ee 
and the ratio of viscous to thermal dissipation as given by the Prandtl number
\be
\Pra \equiv \nu/\kappa\ .
\label{eq:Pr}
\ee 
Here $\alpha$ is the isobaric thermal expansion coefficient, $g$ the gravitational acceleration, $\kappa$ the thermal diffusivity, $\nu$ the kinematic viscosity, and $\Delta T \equiv T_b - T_t$ the applied temperature difference between the bottom ($T_b$) and the top ($T_t$) plate. 

In the present paper we present measurements of the heat transport in the form of the scaled effective thermal conductivity known as the Nusselt number, which is given by
\be
\Nu \equiv \frac{Q L}{A \Delta T \lambda}\ .
\label{eq:Nu}
\ee
Here $Q$ is the applied heat current, $A = D^2\pi/4$ the sample cross-sectional area, and $\lambda$ the thermal conductivity. The measurements cover the range  $5\times 10^{11} \alt \Ra \alt 2\times 10^{14}$ and are for \Pra\ ranging from 0.79 at the lowest to 0.86 at the highest \Ra. 

All fluid properties needed to calculate \Ra, \Pra, and \Nu\ were evaluated at the mean temperature $T_m = (T_t+T_b)/2$ of the sample. They were obtained from numerous papers in the literature, as discussed in Ref.~\cite{LA97}. A small correction for the nonlinear contribution of the side-wall conductance \cite{Ah00,RCCHS01} to the heat carried by the sample was no more that 3\% and was applied to the data.  

\begin{figure}
\centerline{\includegraphics[width=4in]{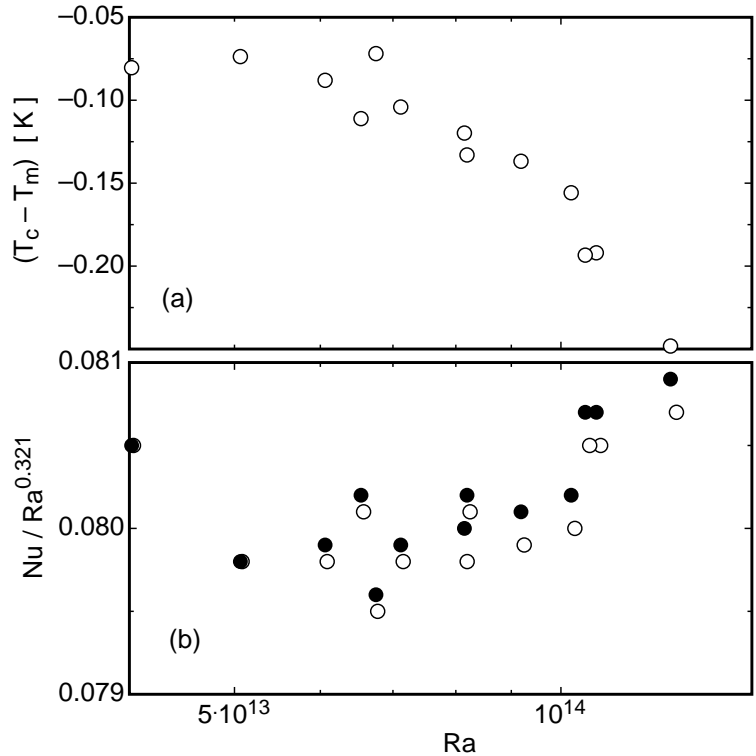}}
\caption{(a): The difference between the center temperature $T_c$ and the mean temperature $T_m$ of the sample as a function of $\Ra$. The sample pressure was  18.6 bars. (b): The reduced Nusselt number $\Nu/\Ra^{0.321}$ as a function of the Rayleigh number \Ra. Solid circles: all fluid properties evaluated at $T_m$. Open circles: All fluid properties evaluated at $T_c$.}
\label{fig:Tm_vs_Tc}
\end{figure}

In a recent communication \cite{UHKMSS12} it was suggested that the fluid properties should be evaluated at the sample center temperature $T_c$ rather than at $T_m$ in order to avoid or minimize effects due to departures from the Oberbeck-Boussinesq (OB) approximation \cite{OB79,Bo03}. We note that this would be contrary to the convention adopted in the usual studies of non-OB effects (see, for instance, \cite{ABFFGL06,AFFGL07,ACFFGLS08}). Nonetheless we explored the importance of the choice between $T_c$ and $T_m$ for our data. In Fig.~\ref{fig:Tm_vs_Tc}a we show $T_c - T_m$ as a function of \Ra\ at the largest \Ra\ of our work where its magnitude is also largest. In Fig.~\ref{fig:Tm_vs_Tc}b we show the corresponding reduced Nusselt numbers $\Nu/\Ra^{0.321}$ as a function of \Ra. The solid (open) circles are based on fluid properties evaluated at $T_m$ ($T_c$). One sees that the largest difference, which occurs at the largest \Ra, is only about a third of a percent. Such a difference is essentially negligible and does not influence the interpretation of our results.

The data obtained in this study are presented as an Appendix to this paper.

\section{Apparatus}
\label{sec:apparatus}

\begin{figure}
\centerline{\includegraphics[width=4in]{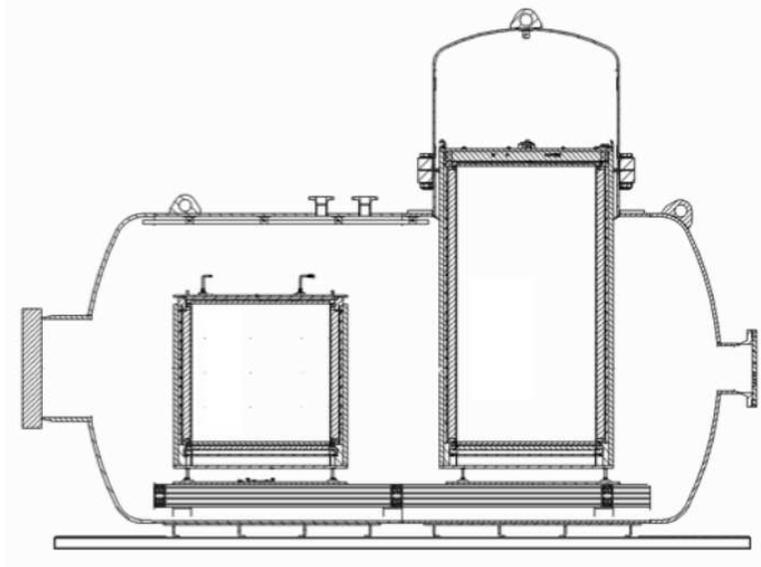}}
\caption{A schematic diagram, to scale, of the Uboot with HPCF-II (right) and HPCF-IV (left).}
\label{fig:Uboot}
\end{figure}

The apparatus was the same as the one described before \cite{AFB09}, except that a new sample cell, known as the High-Pressure Convection-Facility IV or HPCF-IV, was constructed. This cell had an internal height $L = 1120 \pm 2$ mm and a diameter $D$ equal to $L$, yielding an aspect ratio $\Gamma \equiv D/L = 1.000 \pm 0.004$. It had the aluminum top and bottom plates described in Ref.~\cite{AFB09},  and a 9.5 mm thick plexiglas side wall. The plates were sealed to the side wall, and a tube of 13 mm diameter entered the HPCF-IV at mid height through the side wall to permit filling the sample cell with gas to the desired pressure. This tube was sealed by a remotely controlled valve after the sample was filled and all transients had decayed, yielding a completely closed sample. All thermal shields were duplicates of those used for another sample with $\Gamma = 0.50$ known as HPCF-II \cite{FBA09}, except that the side shield was of course shorter. 
The HPCF-IV was located in a high-pressure vessel known as the Uboot of G\"ottingen which could be filled with sulfur hexafluoride (SF$_6$) at pressures up to 19 bars. The Uboot could contain HPCF-IV and as well as HPCF-II simultaneously, as shown in the schematic diagram Fig.~\ref{fig:Uboot}. Completely separate instrumentation and temperature-controlled water circuits enabled simultaneous measurements in the two units. We refer to our previous publications \cite{AFB09,AHFB12} for detailed descriptions of all construction details and experimental procedures.

\section{Results}
\label{sec:Results}

\subsection{Classical state}

\begin{figure}
\centerline{\includegraphics[width=4in]{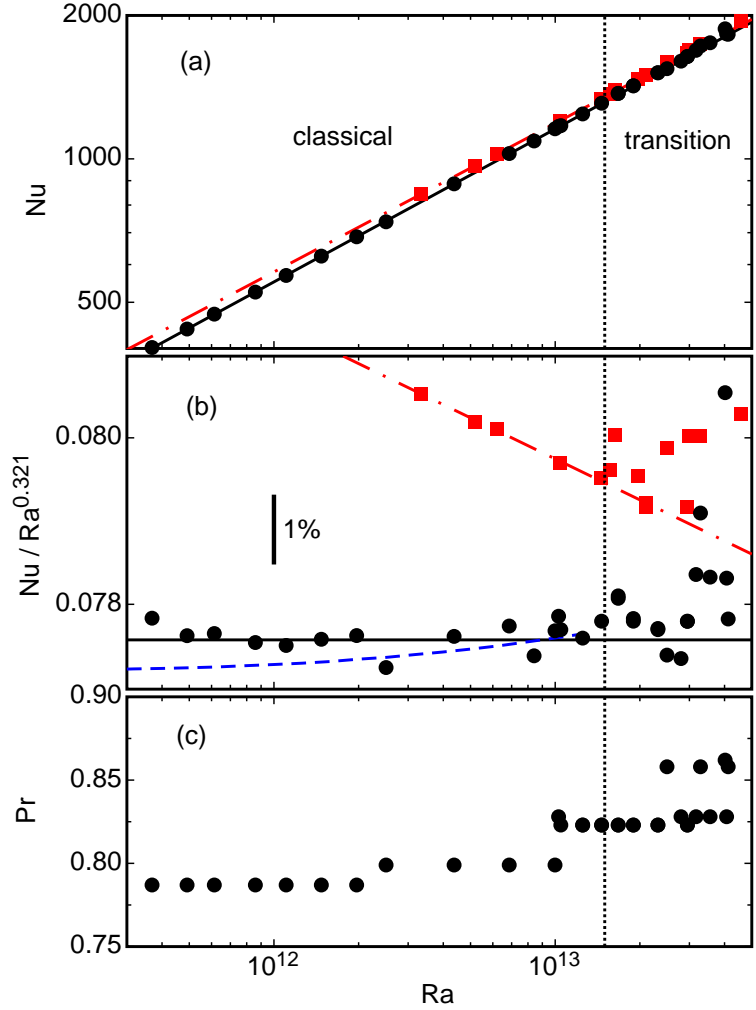}}
\caption{(a): The Nusselt number \Nu\ as a function of the Rayleigh number \Ra\ on logarithmic scales. Black circles: $\Gamma = 1.00$ (HPCF-IV, current work). Red squares: $\Gamma = 0.50$ (HPCF-IIe, Refs.~\cite{HFNBA12,AHFB12}).  The solid black line is a power-law fit to the data for $\Gamma = 1.00$ in the classical state $\Ra < 1.5\times 10^{13}$. The fit gave $N_0 = 0.0764\pm 0.0015$ and $\gamma_{eff} = 0.3216\pm 0.0007$. The red dash-dotted line is the power-law fit to the data in the classical state for $\Gamma = 0.50$ which gave $N_0 = 0.1404, \gamma_{eff} = 0.312$, see Ref.~\cite{HFNBA12,AHFB12}. The vertical dotted line is the location of $\Ra_1^*$ as determined for $\Gamma = 0.50$ \cite{HFNBA12,AHFB12}. (b): The reduced Nusselt number $\Nu/\Ra^{0.321}$ as a function of \Ra\ on logarithmic scales. All symbols and lines are as in (a). We added the blue short-dashed line, which is the prediction of Grossmann and Lohse \cite{GL01}. (c): The Prandtl numbers as a function of \Ra\ for all data points within the range of this figure.}
\label{fig:Nu_of_Ra}
\end{figure}

In Fig.~\ref{fig:Nu_of_Ra}a we show results for \Nu\ as a function of \Ra\ on double logarithmic scales. The data for $\Gamma = 1.00$ are shown in black, and previously published results \cite{HFNBA12,AHFB12} for $\Gamma = 0.50$ (HPCF-II) are given in red. One sees that, within the resolution of this graph, there is very little difference between the data for the two $\Gamma$ values. Also shown in this figure, as a vertical dotted line, is the approximate upper limit of the classical regime and the beginning of the transition range to the ultimate state at $\Ra_1^* = 1.5\times 10^{13}$ as determined from the $\Gamma = 0.50$ data and reported in Ref.~\cite{HFNBA12}. 

The solid black and dash-dotted red lines are fits of the power law
\be
\Nu = N_0 \Ra^{\gamma_{eff}}
\ee
 to the data with $\Ra < \Ra_1^*$. As reported elsewhere \cite{AHFB12}, the fit to the $\Gamma = 0.50$ data yielded $\gamma_{eff} = 0.312 \pm 0.002$. The fit to the $\Gamma = 1.00$ data for $\Ra < \Ra_1^*$ gave $N_0 = 0.0764 \pm 0.0015$ and $\gamma_{eff} = 0.3216 \pm 0.0007$ where the uncertainties are the standard errors of the parameters. The average value of \Pra\ over the range of the data used in the fit was 0.80. Additional possible systematic errors, primarily due to uncertainties in the side-wall correction, lead us to the best estimate $\gamma_{eff} = 0.321 \pm 0.002$ for the Nusselt exponent for $\Gamma = 1.00$ and $\Pra = 0.80$. Fixing  $\gamma_{eff}$ at the value  0.321 let to the amplitude $N_0 = 0.0776$.

In order to provide a better comparison of these two data sets, we show the results in the form of the reduced Nusselt number $\Nu/\Ra^{0.321}$ as a function of \Ra\ on double logarithmic scales in Fig.~\ref{fig:Nu_of_Ra}b. Now the $\Gamma = 1.00$ data scatter about the horizontal solid black line, with the scatter corresponding to a standard deviation of   0.21\%. 

In Fig.~\ref{fig:Nu_of_Ra}b the power-law fit to the $\Gamma = 0.50$ data is shown again as a red dash-dotted line. One can readily see the positive deviations and enhanced scatter of the $\Gamma = 0.50$ data for $\Ra > \Ra_1^*$ where the transition to the ultimate state is beginning. We note that the enhanced scatter is not due to a sudden increase in experimental scatter, but rather a reflection of the intrinsic irreproducibility of the state of the system. Remarkably, also the $\Gamma = 1.00$ data begin to show positive deviations from the horizontal black line and enhanced scatter, suggesting that also the $\Gamma = 1.00$ system is undergoing a similar transition to the ultimate state, beginning at about the same $\Ra_1^*$ that was found for $\Gamma = 0.50$. We shall return to that issue below in Sec.~\ref{sec:transition}.

We also show in Fig.~\ref{fig:Nu_of_Ra}b, as a short-dashed blue line, the prediction of Grossmann and Lohse \cite{GL01} (GL) for $\Nu(\Ra)$ in the classical state with $\Pra = 0.80$. This prediction is based on two coupled equations with several parameters which had been determined by fits to experimental data \cite{AX01} for $\Gamma = 1.00$ over the parameter ranges $4 \alt \Pra \alt 34$ and $3\times 10^7 \alt \Ra \alt 2\times 10^9$. One sees that the comparison with the present data up to $\Ra = 10^{13}$ and for $\Pra \simeq 0.80$ requires a considerable extrapolation. Thus, the excellent agreement is indeed remarkable. Not only does it require a high degree of reliability of the GL equations; it also requires excellent consistency between the experimental data used to determine the free parameters in these equations and the present data. 

\subsection{Comparison with published data}

\begin{figure}
\centerline{\includegraphics[width=4in]{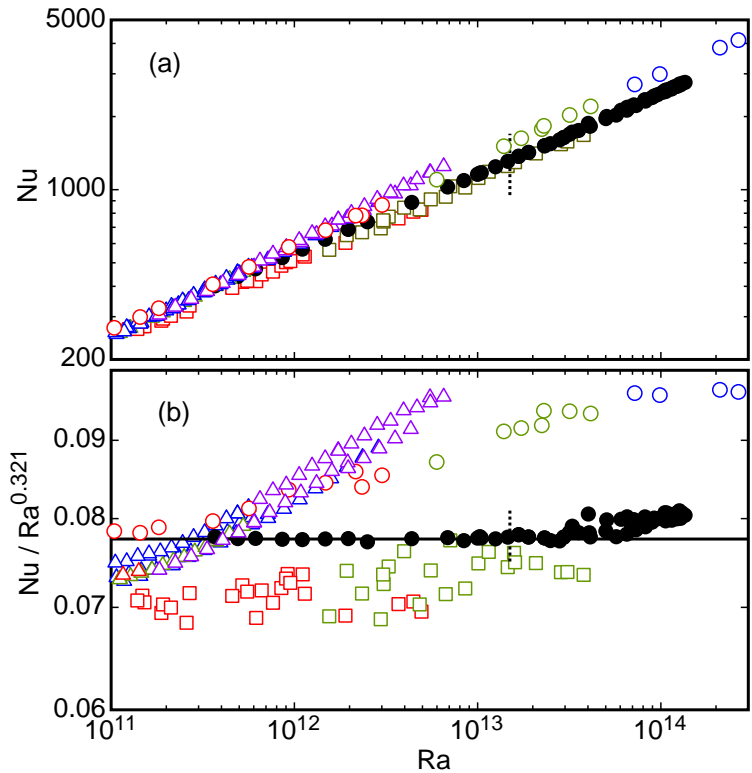}}
\caption{Comparison of our results (black solid circles, $0.79 \leq \Pra \leq 0.86$) with the data of previous investigations. For the previous work we show data with $\Pra < 1$ in red, data with $1 < \Pra < 2$ in green, data with $2 < \Pra < 4$ in blue, and data with $\Pra > 4$ in purple. Open circles: Niemela and Sreenivasan \cite{NS03}. Open up-pointing triangles: Roche {\it et al.} \cite{RGKS10}. Open squares: Urban {\it et al.} \cite{UMS11}. The short vertical dotted line represents $\Ra_1^* = 1.5\times 10^{13}$ as determined for $\Gamma = 0.50$ \cite{HFNBA12,AHFB12}.}
\label{fig:compare}
\end{figure}

In Fig.~\ref{fig:compare} we compare our results with other published data for $\Gamma = 1.00$ (a detailed comparison with literature data for $\Gamma = 0.50$ is being presented elsewhere \cite{AHFB12}). Here too we show \Nu\ in Fig.~\ref{fig:compare}a and, for higher resolution, $\Nu/\Ra^{0.321}$ in  Fig.~\ref{fig:compare}b. For the literature data we use red symbols for data with $\Pra < 1$, green symbols for data with $1 < \Pra < 2$, blue symbols for data with $2 < \Pra < 4$, and purple symbols for data with $\Pra > 4$. Our own data span the range from $\Pra = 0.79$ at the lowest to $\Pra = 0.86$ at the highest \Ra.

The present results are given as solid black circles. The data of Niemela and Sreenivasan \cite{NS03} are shown as open circles. For \Ra\ near $10^{11}$ they follow a power law with an exponent near 0.33; but for $3\times 10^{11} \alt \Ra \alt 5\times 10^{13}$ they rise more steeply, only to level off again for larger \Ra\ to a dependence describable once more by an effective exponent near 0.33. This behavior was attributed by the authors \cite{NS10} to a special type of non-Boussinesq effect near critical points. Thus the data do not yield reliable parameters of a power law for $\Nu(\Ra)$ in the classical region that could be compared with the prediction of GL \cite{GL01}; according to the authors \cite{NS10} the data also do not yield any evidence for a transition to the ultimate state. 

The data for the ``short cell" of Roche {\it et al.} are shown as open up-pointing triangles. They reveal a gradual increase of an effective exponent, starting near $\Ra = 5\times 10^{10}$. Although the authors believe that this rise of the exponent is indicative of an ultimate-state transition at $\Ra_U \simeq10^{11}$, we do not find the evidence convincing. Particularly troublesome  is the low value of $\Ra_U$; it is unlikely that  the boundary-layer shear Reynolds-number $Re_s$ can be high enough to drive the BLs turbulent at so low a value of \Ra\ \cite{HFNBA12}. Very recent direct numerical simulations (DNS) for $\Gamma = 1$ and $\Pra = 0.7$ \cite{WSW12} suggest that $Re_s \simeq 65$ for $\Ra = 10^{11}$, a value much too low to expect a shear instability to turbulence (for the higher \Pra\ values of the experiment $\Rey_s$ would be even lower). On  the other hand, we do not have an alternative explanation of the rise of $\gamma_{eff}$ indicated by these data.

 A third set of data (open squares in  Fig.~\ref{fig:compare}) was published recently by Urban {\it et al.} \cite{UMS11}. They extend up to $\Ra \simeq 4\times 10^{13}$. Although at constant $\Pra \simeq 0.8$ one might have hoped to have reached $\Ra_1^*$ at that point, \Pra\  also rose significantly at these large \Ra, and one expects that $\Ra^*$ increases significantly with \Pra. In any event, no indication of an ultimate-state transition is seen in these data, nor is one claimed by their authors.
 
In view of the above it is our view that the ultimate-state transition has not yet been seen in any of the published data for $\Gamma \simeq 1$.

\subsection{Transition toward the ultimate state}
\label{sec:transition}

\begin{figure}
\centerline{\includegraphics[width=4in]{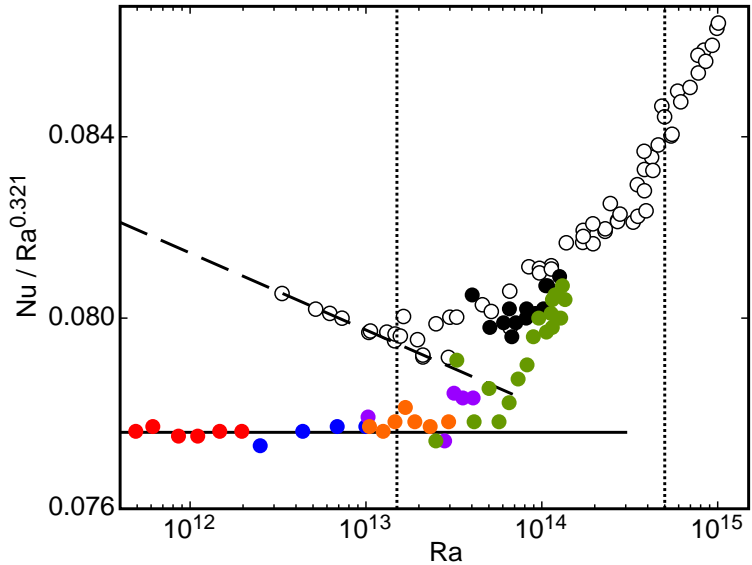}}
\caption{High-resolution comparison of the results for $\Gamma = 0.50$ (HPCF-IIe, Ref.~\cite{HFNBA12,AHFB12}, open symbols) with the present results for $\Gamma = 1.00$ (HPCF-IV, solid circles). For $\Gamma = 1.00$ the data were taken at pressures of 4.1 bars (red, $\Pra = 0.79$), 7.9 bars (blue, $\Pra = 0.80$), 12.1 bars (orange, $\Pra = 0.82$), 12.9 bars (purple, $\Pra = 0.83$), 17.7 bars (green, $\Pra = 0.86$), and 18.6 bars (black, $\Pra = 0.86$). The vertical dotted lines are the locations $\Ra_1^* = 1.5\times 10^{13}$ and $\Ra_2^* = 5\times 10^{14}$ as determined for $\Gamma = 0.50$. The dashed line is a power-law fit to the data in the classical state for $\Gamma = 0.50$ and corresponds to $N_0 = 0.1044$ and $\gamma_{eff} = 0.312$. The solid line is a power-law fit to the data in the classical state for $\Gamma = 1.00$ (HPCF-IV) which gave $N_0 = 0.0776$ and $\gamma_{eff} = 0.321$.}
\label{fig:transition}
\end{figure}

Recent measurements \cite{HFNBA12}  for a $\Gamma = 0.50$ sample revealed that the transition  to the ultimate state for that aspect ratio occurred over the approximate range from $\Ra_1^* = 1.5\times 10^{13}$ to $\Ra_2^* = 5\times 10^{14}$. We show those data in Fig.~\ref{fig:transition} as open circles and compare them with our new data for $\Gamma = 1.00$ (solid circles). The vertical dotted lines in the figure indicate the locations of $\Ra_1^*$ and $\Ra_2^*$. In the classical range below $\Ra_1^*$ both data sets follow a power law, albeit with the slightly different exponents of 0.312 for $\Gamma = 0.50$ and 0.321 for $\Gamma = 1.00$. Near $\Ra_1^*$ {\it both} data sets rise above their respective classical-state power laws, and the enhanced scatter of both  data sets reveals the intrinsic irreproducibility of the state of the system in the \Ra\ range of the transition from the classical to the ultimate state. Although in the classical state the two systems had slightly different Nusselt numbers, it is remarkable that in the transition region they display the same \Nu\ values within the resolution allowed by the intrinsic scatter of the two systems. Unfortunately, in view of the smaller height of the $\Gamma = 1.00$ sample, our measurements are limited to $\Ra \alt 2\times 10^{14}$. Thus it is not possible for us to follow the transition all the way beyond $\Ra_2^*$, as was done for $\Gamma = 0.50$ with HPCF-II.

Finally we note that an extrapolation of the shear Reynolds numbers obtained from DNS \cite{WSW12} for $\Gamma = 1.00$ and $\Pra = 0.7$ yields $\Rey_s \simeq 250$ for $\Ra = \Ra_1^* = 1.5\times 10^{13}$. This is a reasonable value for the onset of the boundary-layer shear transition to the ultimate state. It is also similar to the value of $\Rey_s(\Ra_1^*)$ deduced from experimental determinations of \Rey\ for $\Gamma = 0.50$ \cite{HFNBA12}.

\section{Summary}
\label{sec:summary}

In this paper we presented new data for heat transport, expressed as the Nusselt number \Nu,  by turbulent Rayleigh-B\'enard convection in a cylindrical sample of aspect ratio $\Gamma = 1.00$ over the \Ra\ range $4\times 10^{11} \alt \Ra \alt 2\times 10^{14}$. We note that the Prandtl number was nearly constant for our work, varying only from about 0.79 at our smallest to about 0.86 at our largest \Ra. This stands in contrast to other measurements \cite{NS03,RGKS10,UMS11} which were made near the critical point of helium, where \Pra\ typically varied from about 0.7 to about 4 over the same \Ra\ range. Maintaining a constant \Pra\ is important in the search for the ultimate-state transition because the transition range is expected to shift to larger \Ra\ as \Pra\ increases,  approximately in proportion to $\Pra^{1.6}$ \cite{GL02}.

In the classical regime for Rayleigh numbers $\Ra \alt \Ra_1^* = 1.5\times 10^{13}$ we found that our measurements are in remarkably good agreement with the predictions of Grossmann and Lohse \cite{GL01} (GL). We note that this agreement not only implies excellent reliability of the prediction. It also indicates consistency of the new data for $\Pra \simeq 0.8$ and $5\times 10^{11} \alt \Ra \alt 1.5\times 10^{13}$ with measurements \cite{AX01}  made a decade ago, using very different experimental techniques and organic fluids rather than compressed gases, since these older data for $4 \alt \Pra \alt 34$ and $3\times 10^7 \alt \Ra \alt 2\times 10^9$ were used to fix the free parameters of the equations derived by GL.  

We compared the $\Gamma = 1.00$ results with previous measurements for $\Gamma = 0.50$. In the classical regime we found that the two geometries yielded slightly different effective exponents of the power laws that describe $\Nu(\Ra)$. For $\Gamma = 0.50$ we reported elsewhere \cite{AHFB12} that $\gamma_{eff} = 0.312 \pm 0.002$. For $\Gamma = 1.00$ we now find that $\gamma_{eff} = 0.321 \pm 0.002$, in excellent agreement with the GL result $\gamma_{eff} = 0.323$ in our parameter range.

In the classical range $\Ra \alt \Ra_1^* = 1.5\times 10^{13}$ the data had very little scatter, with root-mean-square deviations from the power-law fit as small at 0.2\%. At larger \Ra\ the scatter increased, indicating an intrinsic irreproducibility of the state of the system from one data point to another. Further, most of the points for  $\Ra > \Ra_1^*$ fell well above the power-law extrapolation from the classical state. Both of these phenomena were seen as well at the beginning of the transition to the ultimate state for $\Gamma = 0.5$ \cite{HFNBA12}. Indeed, for $\Ra > \Ra_1^*$ the $\Gamma = 1.00$ data agree quite closely with the $\Gamma = 0.50$ data. Thus we believe that we observed the onset of the transition to the ultimate state also for $\Gamma = 1.00$, and that $\Ra_1^*$ for $\Gamma = 1.00$ is very nearly the same as it is for $\Gamma = 0.50$.  Earlier measurements by Roche {\it et al.} \cite{RGKS10} had revealed a transition in $\Nu(\Ra)$ at several $\Gamma$ values at  Rayleigh numbers $\Ra_U \simeq 1.3\times 10^{11} \Gamma^{-2.5\pm 0.5}$ which those authors attributed to the ultimate-state transition (for a detailed discussion of some of those data, see Ref.~\cite{AHFB12}). In contradistinction to this result, the transitions found by us for $\Gamma = 0.50$ and 1.00 are, within the resolution of the data,  independent of $\Gamma$. We believe that a $\Gamma$-independent $\Ra^*_1$ suggests that the boundary-layer shear-transition is induced by fluctuations on a scale less than the sample dimensions rather than by a global $\Gamma$-dependent flow mode.  Above $\Ra_1^*$ any difference between the heat transport for the two $\Gamma$ values is too small to be resolved, suggesting a universal aspect of the ultimate-state transition and properties. 
Unfortunately the smaller height of the $\Gamma = 1.00$ sample, compared to $\Gamma = 0.50$, limits the accessible range to $\Ra \alt 2\times 10^{14}$. Thus, for this case,  we were able to cover only a little more than the lower half of the transition range to the ultimate state.

\ack

We are very grateful to the Max-Planck-Society and the Volkswagen Stiftung, whose generous support made the establishment of the facility and the experiments possible. We thank the Deutsche Forschungsgemeinschaft (DFG) for financial support through SFB963: ``Astrophysical Flow Instabilities and Turbulence".  The work of G.A. was supported in part by the U.S National Science Foundation through Grant DMR11-58514. We thank Andreas Kopp, Artur Kubitzek, and Andreas Renner for their enthusiastic technical support. We are very grateful to Holger Nobach for many useful discussions and for his contributions to the assembly of the experiment.

\vfill\eject

\appendix

\section{Data tables.}

 \begin{table}[h]
\caption{SF$_6$, HPCF-IV. The data are presented in chronological order.}
\vskip 0.1in
\begin{center}
\begin{tabular}{ccccccc}

Run No. &	 $P$ (bars) &	$T_m (^\circ{\mathrm C})$ &	$\Delta T$ (K) &	\Ra &	\Pra &	\Nu\\
\hline
120227&	18.557&	21.067&	6.534&	6.752e+13&	0.862&	2189.32\\
120228&	18.540&	21.309&	10.603&	1.078e+14&	0.862&	2577.65\\
120229&	18.537&	21.551&	10.492&	1.053e+14&	0.862&	2559.41\\
120301&	18.571&	21.536&	12.461&	1.262e+14&	0.862&	2720.43\\
120302&	18.581&	21.599&	8.091&	8.192e+13&	0.862&	2346.32\\
120303&	18.555&	21.546&	6.486&	6.541e+13&	0.862&	2183.10\\
120304&	18.530&	21.557&	4.012&	4.019e+13&	0.862&	1874.97\\
120307&	18.559&	21.503&	5.005&	5.063e+13&	0.862&	2001.59\\
120308&	18.566&	21.491&	5.978&	6.061e+13&	0.862&	2122.51\\
120309&	18.574&	21.508&	7.011&	7.117e+13&	0.862&	2233.55\\
120310&	18.583&	21.506&	8.007&	8.147e+13&	0.862&	2335.39\\
120311&	18.592&	21.509&	9.011&	9.187e+13&	0.862&	2429.87\\
120312&	18.600&	21.502&	9.996&	1.022e+14&	0.862&	2517.33\\
120314&	4.058&	21.508&	4.020&	4.910e+11&	0.787&	439.40\\
120315&	4.068&	21.517&	16.029&	1.968e+12&	0.787&	686.14\\
120316&	4.064&	21.518&	12.035&	1.474e+12&	0.787&	624.99\\
120317&	4.061&	21.516&	9.032&	1.105e+12&	0.787&	569.24\\
120318&	4.060&	21.513&	7.027&	8.589e+11&	0.787&	525.25\\
120319&	4.058&	21.513&	5.026&	6.137e+11&	0.787&	472.17\\
120320&	4.057&	21.511&	3.021&	3.686e+11&	0.787&	401.83\\
120322&	7.953&	21.499&	15.992&	9.984e+12&	0.799&	1156.44\\
120323&	7.944&	21.511&	11.030&	6.863e+12&	0.799&	1026.08\\
120324&	7.940&	21.518&	7.033&	4.370e+12&	0.799&	886.28\\
120325&	7.934&	21.520&	4.038&	2.504e+12&	0.799&	737.70\\
120329&	12.919&	21.515&	4.027&	1.028e+13&	0.828&	1169.92\\
120331&	12.894&	21.524&	11.043&	2.801e+13&	0.828&	1603.61\\
120401&	12.913&	21.475&	15.941&	4.068e+13&	0.828&	1829.93\\
120402&	12.903&	21.485&	13.963&	3.554e+13&	0.828&	1752.55\\
120403&	12.898&	21.485&	12.464&	3.167e+13&	0.828&	1689.55\\
120404&	17.649&	21.409&	11.808&	9.604e+13&	0.858&	2462.53\\
120405&	17.612&	21.357&	16.199&	1.309e+14&	0.858&	2742.66\\
120406&	17.528&	20.463&	14.413&	1.188e+14&	0.857&	2655.13\\
120407&	17.533&	20.472&	13.931&	1.149e+14&	0.858&	2623.51\\
120408&	17.616&	20.454&	13.396&	1.129e+14&	0.858&	2596.70\\
120409&	17.689&	21.454&	16.534&	1.355e+14&	0.858&	2764.66\\

\end{tabular}
\end{center}
\label{tab:HPCF}
\end{table}
\vfill\eject

 \begin{table}[h]
\begin{center}
\begin{tabular}{ccccccc}

120410&	17.737&	21.493&	15.472&	1.281e+14&	0.859&	2702.71\\
120411&	17.727&	21.411&	14.807&	1.227e+14&	0.859&	2663.98\\
120413&	17.683&	21.513&	4.025&	3.285e+13&	0.858&	1725.56\\
120415&	17.763&	21.439&	9.869&	8.241e+13&	0.859&	2316.20\\
120416&	17.725&	21.544&	6.086&	5.011e+13&	0.859&	1961.69\\
120417&	17.741&	21.392&	13.774&	1.147e+14&	0.859&	2601.23\\
120417&	17.730&	21.407&	12.802&	1.062e+14&	0.859&	2535.08\\
120418&	17.710&	21.429&	10.850&	8.949e+13&	0.859&	2394.85\\
120418&	17.689&	21.480&	8.954&	7.330e+13&	0.858&	2220.76\\
120419&	17.678&	21.502&	6.999&	5.709e+13&	0.858&	2028.18\\
120419&	17.674&	21.501&	7.997&	6.517e+13&	0.858&	2124.46\\
120420&	17.679&	21.531&	5.057&	4.121e+13&	0.858&	1826.17\\
120420&	17.678&	21.535&	3.066&	2.497e+13&	0.858&	1546.36\\
120423&	12.173&	21.497&	13.987&	2.951e+13&	0.823&	1639.95\\
120424&	12.160&	21.517&	11.031&	2.318e+13&	0.823&	1516.00\\
120425&	12.149&	21.527&	9.052&	1.896e+13&	0.823&	1423.45\\
120426&	12.131&	21.518&	8.032&	1.675e+13&	0.823&	1372.14\\
120427&	12.097&	21.539&	7.077&	1.462e+13&	0.823&	1308.59\\
120428&	12.096&	21.531&	6.060&	1.252e+13&	0.823&	1242.46\\
120829&	12.088&	21.542&	5.082&	1.047e+13&	0.823&	1174.20\\

\end{tabular}
\end{center}
\label{tab:HPCF}
\end{table}

\vfill
\eject


\end{document}